\documentclass[reprint,amsmath,amssymb,aps]{revtex4-1}
\usepackage{graphicx}
\usepackage{dcolumn}
\usepackage{bm}
\usepackage{amsmath}
\usepackage{dsfont}
\usepackage{subcaption}
\usepackage{braket}
\usepackage{epstopdf}
 \usepackage{color}	
\begin{document}
\title{Strong entanglement criteria for mixed states, based on uncertainty relations}
\author{Manju}
\email{2018phz0009@iitrpr.ac.in}
\author{Asoka Biswas}
\author{Shubhrangshu Dasgupta}
\affiliation{
Department of Physics, Indian Institute of Technology Ropar, Rupnagar, Punjab 140001, India
}
\date{\today}

\begin{abstract}
We propose an entanglement criterion, specially designed for mixed states, based on uncertainty relation and the Wigner-Yanase skew information. The variances in this uncertainty relation does not involve any classical mixing uncertainty, and thus turns out to be purely of quantum mechanical nature. We show that any mixed entangled state can be characterized by our criterion. We demonstrate its utility for several generalized mixed entangled state including Werner states and it turns out to be stronger than any other known criterion in identifying the correct domain of relevant parameters for entanglement. The proposed criterion reduces to the Schrodinger-Robertson inequality for pure states. 
\end{abstract}
\maketitle
\section{Introduction}
Entanglement between two or more subsystems is considered as a resource to deal with quantum information. Several applications like quantum teleportation, quantum metrology, quantum cryptography, and super-dense coding require that the participating subsystems will be entangled. Identifying whether these subsystems are entangled or not is therefore an essential step toward quantum information processing.

In the past two decades, several criteria for detection of entanglement have been developed. The positive partial transpose (PPT) criterion by Peres and Horodecki \cite{peres1996separability} has been one of the most important ones, which provides necessary and sufficient condition in certain cases. Peres had proposed that the density matrix of a bipartite entangled state after partial transpose (PT) in the basis of one of the parties exhibits negative eigenvalues. The other criteria include those based on reduction \cite{horodecki1999reduction} and the computable cross norm \cite{rudolph2003some}. However, to test these criteria in experiments, one would ideally need to reproduce the density matrix using quantum state tomography. 

As an alternative approach more suitable for experimental detection of entanglement, criteria based on measurement outcomes of the relevant observables have been derived. For example, the PT criterion has been mapped into uncertainty relations of the relevant quadratures, violation of which would indicate existence of entangled states \cite{agarwal2005inseparability,gillet2008multipartite}. 
There exist methods based on Bell-type inequalities \cite{clauser1978bell}, local uncertainty relations \cite{hofmann2003violation},the SRPT inequality \cite{gillet2008multipartite}, as well. These measurement-dependent criteria are often expressed in terms of inequalities, which are satisfied by separable states, and any state violating these inequalities must be entangled. 

These criteria are useful to detect entanglement in pure as well as in mixed states. But unfortunately, they cannot reveal the correct domain of the relevant parameters, as prescribed by the PPT criterion, to detect entanglement in mixed states. This can be attributed to the fact that the mixed states involve both classical and quantum probability distributions and the above criteria do not differentiate between these two for evaluation of the expectation values. As the entanglement is a property purely of quantum nature, to detect this, we need a criterion which considers only the quantum uncertainties of the relevant variables. 

In addition to the criteria based on partial transposition, quantum uncertainty of local observables has also been used to characterize non-classical correlation like quantum discord \cite{girolami2013characterizing}. Quantum discord can distinguish between classical and quantum probability distributions, inherent in the system. However, it is rather quite cumbersome to calculate the discord, as it requires optimization over many measurements and for the systems with more than two qubits, it becomes more intractable. In this work, we use an alternative strategy. 
 
In this paper, we consider a Schrodinger-Robertson-type uncertainty relation proposed by Furuichi \cite{furuichi2010schrodinger}, which includes the Wigner-Yanase skew information. This skew information is known to give a measure of the quantum uncertainty of an operator $X$ with respect to a given state $\rho$. As mentioned above, for a joint state of two subsystems, the quantum uncertainty of a local variable provides an alternative estimate of the discord. Here, we consider both local and non-local variables of the two subsystems and apply the partial transposition criterion to the uncertainty relation to detect entanglement between them. When using the nonlocal variables, we essentially consider the nonlocal correlation of the subsystems. Moreover, the uncertainty relation used for employing the entanglement criterion which involves only the terms which do not contain the classical mixing uncertainty. This implies that, the uncertainty relation is particularly suitable for mixed states. We emphasize that the skew information represents quantum (rather than total) uncertainty which considers both incompatibility and correlation between the relevant observables. Thus the uncertainty relation is of purely quantum nature and is stronger than the conventional ones when one deals with mixed states.  Note that this uncertainty relation reduces to the Schrodinger-Robertson inequality for the pure state.  We will apply the inequality to several generalized mixed states including the Werner states. We show that our inequality reveals an ideal domain of the relevant parameter for entanglement, unlike the other criteria based on the Bell inequalities \cite{bell1964einstein,clauser1969proposed}, uncertainty relations  \cite{guhne2002detection}, and the Schrodinger-Robertson inequality \cite{gillet2008multipartite}. It must be borne in mind that the proposed inseparability inequality requires the full knowledge of the density matrix to be evaluated. So we need quantum state tomography to reproduce the density matrix.    

The paper is organized as follows. In Sec. II, we review some basic properties of skew information and highlight its relation with variance. In Sec. III, We will present the entanglement criterion in the form of the Schrodinger-Robertson type inequality in terms of the skew information. We also demonstrate how the violation of this inequality can detect entanglement for a large class of pure and mixed states, including the Werner states. In Sec. IV, we conclude the paper. 

\section{Wigner-Yanase Skew Information}
In their seminal paper on quantum measurement, Wigner and Yanase introduced the quantity, the skew information, \cite{wigner1997information} as
 \begin{equation}\label{1}
{I(\rho,X)} = -\frac{1}{2} {\rm Tr}[\sqrt{\rho},X]^{2}\;.
\end{equation}
This corresponds to a measure of the amount of information on the values of observable, which is skew to the operator $X$. Here $X$ is a conserved quantity like Hamiltonian, momentum etc., of the relevant quantum system, which is in a state described by the density matrix $\rho$. Note that $I(\rho,X)$ accounts for the non-commutativity between $\rho$ and $X$.

The skew information satisfies several criteria, suitable for a valid  information-theoretic measure, which are as follows:
\begin{enumerate}
\item  Non-negativity: $I(\rho,X) \ge 0$.
\item Convexity: It is convex with respect to $\rho$ in the sense that
\begin{equation}\label{2}
I(p_1\rho_1+p_2\rho_2,X) \le p_1I(\rho_1,X)+p_2I(\rho_2,X)\;,
\end{equation}
where $p_1+p_2 = 1$, $p_1,p_2 \ge 0$. This suggests that the skew information decreases when two density matrices are mixed with each other.
\item Additivity: This is represented by 
\begin{equation}\label{3}
I(\rho_1\otimes\rho_2,X_1\otimes \mathds{I}_2+\mathds{I}_1\otimes X_2)=I(\rho_1,X_1)+I(\rho_2,X_2)\;,
\end{equation}
where $\rho_1$ and $\rho_2$ are two density operators describing the two systems, $\mathds{I}_i$ are the density matrices in their respective basis ($i\in 1,2$), and  $X_1$ and $X_2$ are their corresponding conserved quantities.

\item Let $U$ be a unitary operator, then
\begin{equation}\label{4}
I(U\rho U^{-1},X)=I(\rho,X)\;,
\end{equation}
where, $U=e^{-\iota\theta X}$ commutes with $X$.
\end{enumerate}
i.e, when the state changes according to the Landau-von Neumann equation, the skew information remains constant for isolated systems.

It has been used to construct measures of quantum correlations \cite{luo2012quantifying} and quantum coherence \cite{yu2017quantum}, to detect entanglement \cite{chen2005wigner}, to study phase transitions \cite{lei2016wigner} and uncertainty relations \cite{luo2003wigner,luo2004skew,luo2005heisenberg}, and so on.

Skew information is related to the conventional variance, through the following relation:
\begin{equation}\label{5}
I(\rho,X) = {\rm Tr}(\rho X^{2}) - {\rm Tr}(\sqrt{\rho}X\sqrt{\rho}X)\;. 
\end{equation}
This is equal to the variance only if the state $\rho$ is a pure state, i.e., if $\rho = |\psi\rangle\langle \psi|$:
\begin{equation}\label{6}
I(\rho,X) = V(\rho,X) \;,
\end{equation}
where $V(\rho,X) = {\rm Tr}\rho X^{2} - ({\rm Tr}\rho X)^{2}$.  On the other hand, for any mixed state $\rho$, the skew information is always dominated by the variance:
\begin{equation}\label{7}
I(\rho,X) \le V(\rho,X)\;.   
\end{equation}

A mixed state can be considered as a classical mixture of quantum states. 
The variance does not differentiate between the quantum uncertainty (arising out of purely quantum probability distribution) and the classical uncertainty (associated with the classical mixing) in the mixed state. On the contrary, the skew information can be interpreted as equivalent to quantum uncertainty and does not account for the classical mixing. In fact, it vanishes if $\rho$ and $X$ commute with each other. Also the convexity property $I$, as mentioned above, suggests that classical mixing cannot increase quantum uncertainty. 

The interpretation of the skew information as a kind of quantum uncertainty and the relation (\ref{7}) above was used to construct an uncertainty relation, which is stronger than the usual Heisenberg uncertainty relation to detect entanglement in a mixed state. We will discuss this in the next Section.

\section{Entanglement criteria based on the uncertainty relations}
In this Section, we will first discuss the modified uncertainty relations and then will propose how this can be useful as entanglement criterion. 

\subsection{Modified uncertainty relations}
Usual uncertainty relation, due to Heisenberg, sets a fundamental limit on simultaneous measure of two non-commuting observables \cite{holevo2011probabilistic}. For measurement of any two observables $A$ and $B$ in a quantum state $\rho$, this is given by
\begin{equation}\label{8}
V(\rho,A)V(\rho,B)\ge \frac{1}{4}|{\rm Tr}(\rho[A,B])|^{2}\;,
\end{equation}
where 
 $V(\rho,A)$ and $V(\rho,B)$ are  the variances of $A$ and $B$, as defined above, and ${\rm Tr}(\rho[A,B])$ is the average of commutator $[A,B] = AB - BA$ in the state $\rho$.
It is noticeable that the commutator, which is so fundamental in quantum mechanics, makes its appearance in Heisenberg's relation. In addition to this commutator, one also considers the correlation between the observables, which is usually expressed in terms of anti-commutator in quantum mechanics. This was included by  Schrodinger \cite{schrodinger1999heisenberg} into the canonical form of the uncertainty relation, that now takes the following form: 
\begin{equation}\label{10}
V(\rho,A)V(\rho,B)\ge \frac{1}{4}|{\rm Tr}(\rho[A,B])|^{2}+ \frac{1}{4}|{\rm Tr}(\rho\{A_0,B_0\})|^{2}\;.
\end{equation}
Here ${\rm Tr}[\rho\{A_0,B_0\}]$ denotes the average of the anti-commutator $\{A_0,B_0\}=A_0B_0+B_0A_0$, where $A_0 = A - \langle A\rangle_\rho \mathds{I}$ and
$B_0 = B - \langle B\rangle_\rho \mathds{I}$ can be interpreted as the fluctuation operators about their respective expectation values, calculated for the state $\rho$. 

 %



As discussed in the Sec. II, the skew information can be considered as quantum uncertainty. Luo therefore proposed \cite{luo2003wigner} that Heisenberg's uncertainty relation might be changed, as follows, in terms of the skew information, for any two observables $A,B$ and the quantum state $\rho$, 
\begin{equation}\label{13}
I(\rho,A)I(\rho,B) \ge \frac{1}{4}|{\rm Tr}(\rho[A,B])|^{2}\;.
\end{equation}
This relation is defined in the spirit of the relation $0 \le \ I(\rho,A) \le V(\rho,A)$. However, this does not distill the right essence of the uncertainty relation, as when the quantum uncertainties $I$ vanish for two non-commuting operators $A$ and $B$, the above inequality $(\ref{13})$ gets violated, even if the state $\rho$ has non-classical correlations. 

It was later observed that the Heisenberg uncertainty relation is of purely quantum nature for pure state and is of "mixed" flavor for mixed state because $V(\rho,A)$ is a hybrid of classical and quantum uncertainty for these state. Motivated by this simple observation, Luo then introduced \cite{luo2005heisenberg} the quantity $U(\rho,A)$, as follows, by decomposing the variance into classical and quantum parts i.e., $V(\rho,A) = C(\rho,A) + I(\rho,A)$:
\begin{eqnarray}\label{14}
U(\rho,A)&=& \sqrt{V^{2}(\rho,A) - C^{2}(\rho,A)}\nonumber\\
&=&\sqrt{V^{2}(\rho,A) - [V(\rho,A)-I(\rho,A)]^{2}}\;.
\end{eqnarray}
Luo then successfully introduced a new Heisenberg-type uncertainty relation based on $U(\rho,A)$ (which suitably takes care of exclusion of classical mixing, specially for mixed state) as follows:
\begin{equation}\label{15}
U(\rho,A)U(\rho,B) \ge \frac{1}{4}|{\rm Tr}(\rho[A,B])|^{2}\;.
\end{equation}
The three quantities $V(\rho,A)$, $I(\rho,A)$, and $U(\rho,A)$ have the following ordering:
\begin{equation}\label{16}
0 \le I(\rho,A) \le U(\rho,A) \le V(\rho,A)\;.
\end{equation}
Clearly, for pure states, we have the classical correlation $C=0$ and thus, $U=V$ and the above relation (\ref{15}) becomes the same as the original uncertainty relation (\ref{8}).

The above uncertainty relation (\ref{15}) is improved by Furuichi \cite{furuichi2010schrodinger}, who proposed a stronger Schrodinger-type uncertainty relation, by improving upper bound, for the quantity $U$, as
\begin{equation}\label{17}
U(\rho,A)U(\rho,B) - |{\rm Re}\{C_\rho(A,B)\}|^{2} \ge \frac{1}{4} |{\rm Tr}(\rho [A,B])|^{2}\;,
\end{equation}
where $C_\rho(A,B)$ is called Wigner-Yanase correlation between two observables and can be written as
\begin{equation}\label{11}
C_\rho (A,B) = {\rm Tr}(\rho A^{*}B) - {\rm Tr}(\sqrt{\rho}A^{*}\sqrt{\rho}B)\;,
\end{equation}
where $A^*$ is the complex conjugate of the operator $A$. Note that, if $A=B$ are self-adjoint,  this simplifies to 
\begin{equation}\label{12}
C_\rho(A,A) = {\rm Tr}(\rho A^{2}) - {\rm Tr}(\sqrt{\rho}A\sqrt{\rho}A)\;,
\end{equation}
which becomes the same as the skew information.
It can be shown that
\begin{equation}\label{18}
|{\rm Im}\{C_\rho(A,B)\}|^{2} = \frac{1}{4} |{\rm Tr}(\rho[A,B])|^{2}\;.
\end{equation}
So the inequality (\ref{17}) can be finally written as 
\begin{equation}\label{19}
U(\rho,A)U(\rho,B) \ge |C_\rho(A,B)|^{2}\;.
\end{equation}

\subsection{Relation to the entanglement criteria}
As mentioned in the Introduction, entanglement criteria based on partial transpose in the uncertainty relations do not differentiate between the quantum and classical probabilities and also the correlations of the observables. Thus they fail to attain the correct domain, as prescribed by the PPT criteria, of the relevant variable in the mixed states. As the uncertainty relation (\ref{19}) includes both the quantum uncertainty and correlations, while excluding the classical uncertainty, it is expected that entanglement criterion based on (\ref{19}) would prove to be much stronger compared to the older versions of such criteria, when mixed states are involved.  In the following we therefore propose a new criterion, particularly useful for detecting entanglement in bipartite mixed states. 
\begin{equation}\label{20}
U(\rho^{PT},A)U(\rho^{PT},B) \ge |C_{\rho^{PT}}(A,B)|^{2}\;,
\end{equation}
where $A$ and $B$ are the operators in the joint Hilbert space and $\rho^{PT}$ represents the partial transpose of the joint density matrix $\rho$ in terms of one of the subsystems. Violation of the above inequality is a sufficient condition to entanglement because Peres criterion is sufficient to detect entanglement in bipartite system.

\subsection{Examples}

\subsubsection{Werner state}
To illustrate the utility of the criterion (\ref{20}), we first consider a Werner state,  which is mixture of a maximally entangled state and the maximally mixed state. The Werner state for a two-qubit system is given by 
\begin{equation}\label{21}
\rho = \frac{1-p}{4} \mathds{I}_1\otimes \mathds{I}_2 + p |\psi_-\rangle\langle \psi_-|\;,
\end{equation}
where  $|\psi_-\rangle = \frac{1}{\sqrt{2}}(|01\rangle - |10\rangle)$ is a maximally entangled state (one of the four celebrated Bell states) and  $0 \le p \le 1$. 
In the computational basis $(|00\rangle,|01\rangle,|10\rangle,|11\rangle)$ of two qubits, we can write $\rho$  as follows: 
\begin{equation}\label{22}
\rho = \frac{1}{4}\begin{pmatrix}
1-p & 0 & 0 & 0 \\
0 & 1+p & -2p & 0 \\
0 & -2p & 1+p &0 \\
0 & 0 & 0 & 1-p \\
\end{pmatrix}\;.
\end{equation}
With the partial transpose with respect to the second qubit, this transforms into
\begin{equation}\label{23}
\rho^{PT} = \frac{1}{4}\begin{pmatrix}
1-p & 0 & 0 & -2p \\
0 & 1+p & 0 & 0 \\
0 & 0 & 1+p &0 \\
-2p & 0 & 0 & 1-p \\
\end{pmatrix}\;,
\end{equation}
the eigenvalues of which are $(1+p)/4$ (triply degenerate) and $(1-3p)/4$. 
Clearly, the Werner state is entangled (inseparable) for $p>\frac{1}{3}$ (as one of the eigenvalues becomes negative), according to the PT criterion and maximally entangled when $p=1$. But, if we use the uncertainty relation (\ref{10}) with $\rho^{PT}$, we find the condition for separability as $p\le 1$, which is always satisfied. Thus, the violation of this inequality cannot be reliably used to detect entanglement in this state. If one would perform the partial transposition on the relevant observables A and B, instead of on $\rho$, the uncertainty relation (\ref{10}) becomes the SRPT inequality. However, not all observables are suitable to demonstrate the violation of the SRPT inequality. They have to satisfy a general condition to be eligible. The SRPT inequality detects the entanglement of Werner state when $p>\frac{1}{2}$ for a particular choice of observables A and B. This lower bound is however larger than $p=1/3$. We show below that the present criterion reveals the entanglement even in the domain $(\frac{1}{3}, \frac{1}{2})$.   



As we proposed in this paper, we now explore  the suitability of the uncertainty relation (\ref{20}) for entanglement detection. In this regard, we first obtain
\begin{equation}\label{26}
\sqrt{\rho^{PT}}=\begin{pmatrix}
P & 0 & 0 & Q \\
0 & R & 0 & 0 \\
0 & 0 & R &0 \\
Q & 0 & 0 & P \\
\end{pmatrix}\;,
\end{equation}
where $P=\sqrt{1+p}/2$, $Q=[\sqrt{1-3p} - \sqrt{1+p}]/4$, and $R=[\sqrt{1+p}+\sqrt{1-3p}]/4$. 


It is now important to suitably choose the observables $A$ and $B$, such that they do not commute with $\rho^{PT}$.

{\it Case I:} Following \cite{girolami2013characterizing}, we first choose a set of local observables $A=\sigma_z\otimes \mathds{I}_2$ and $B=\mathds{I}_1\otimes\sigma_z$. For these operators, we found that 
\begin{eqnarray} \label{27}
&&V(\rho^{PT},A)= V(\rho^{PT},B)=1\;,\nonumber\\
&&I(\rho^{PT},A)=I(\rho^{PT},B)=\frac{1}{2}(1-p-\sqrt{1+p}\sqrt{1-3p})\;,\nonumber\\
&&U(\rho^{PT},A)=U(\rho^{PT},B)\nonumber\\
&&=\sqrt{\frac{1}{2}\left\{1+p^{2}-(1+p)^{3/2}\sqrt{1-3p}\right\}}\;,\nonumber\\
&&C_{\rho^{PT}}(A,B)=\frac{1}{2}\left(1- p- \sqrt{1+p}\sqrt{1-3p}\right)\;.
\end{eqnarray}

For $0\le p \le \frac{1}{3}$, the state $\rho^{PT}$ is positive, which implies that it describes some physical state and therefore satisfies the inequality (\ref{20}). For $ p > \frac{1}{3}$, however, the term $\sqrt{1-3p}$ is complex. Therefore, we can rewrite $U(\rho^{PT},A)$ and $U(\rho^{PT},B)$ as
\begin{equation} \label{28}
\sqrt{\frac{1}{2}\left\{1+p^{2}-\iota(1+p)^{3/2}\sqrt{3p-1}\right\}}\;.
\end{equation}



There are several forms of the square root (\ref{28}). By choosing $U(\rho^{PT},A)  = \iota\sqrt{b-a}+\sqrt{a+b}$, and $U(\rho^{PT},B)=\iota\sqrt{b-a}-\sqrt{a+b}$ (where $a = (1+p^{2})/4$ and $b = p\sqrt{p^{2}+2p+2}/2$), and by using (\ref{27}), we have the following condition from (\ref{20}):
\begin{equation} \label{cond1}
p[\sqrt{p^{2}+2p+2}+p]\le 0\;.
\end{equation}
It should be remembered that the above condition is obtained in the domain $p>1/3$ and is obviously violated for all $p\in (1/3,1]$. 

{\it Case II:} Usually, measurement of local observables bypasses the issue of nonlocal correlations that exist between two subsystems, when they are entangled. Accordingly, if we choose a set of global observables $A = \sigma_z\otimes\sigma_z$ and $B = \sigma_x\otimes\sigma_x$, we find that they commute with $\rho^{PT}$, and thus the skew information vanishes, i.e., $I(\rho^{PT},A) = I(\rho^{PT},B) = 0$. The correlation between these operators also vanishes:  $C_{\rho^{PT}}(A,B) = 0$. Clearly, for such choices of global observables, we cannot clearly say anything about the inseparability of the Werner state using (\ref{20}). 

But if we choose a different set of operators, say, $A =\sigma_x\otimes\sigma_y$ and $B=\sigma_y\otimes\sigma_x$, which do not commute with $\rho^{PT}$, we have the same expressions of $V$, $I$, and $U$ as in (\ref{27}), and the correlation term becomes 
\begin{equation}\label{29}
C_{\rho^{PT}}(A,B)=\frac{1}{2}(-1+p +\sqrt{1+p}\sqrt{1-3p})\;.
\end{equation}

Using these expressions, we find that, for $p>\frac{1}{3}$, the condition (\ref{20}) for separability is violated, i.e., the Werner state is entangled for $p>\frac{1}{3}$. This result indicates that the criterion (\ref{20}) is the strongest among all the other known criteria for entanglement. For example, the Bell's inequalities \cite{bell1964einstein,clauser1969proposed} lead to $p>\frac{1}{\sqrt{2}}$ for entanglement, while the uncertainty relation in \cite{guhne2002detection} sets the lower limit as $p>\frac{1}{\sqrt{3}}$ and the Schrodinger-Robertson inequality based on local variables \cite{gillet2008multipartite} suggests $p>\frac{1}{2}$. 

As clear from the two cases discussed above, the separability criterion (\ref{20}) affirms the correct limit for entanglement, as one gets from the Peres criteria. Interestingly, both local and global sets of operators can reveal this limit. One only needs to choose the set of operators that do not commute with the $\rho^{PT}$.

\subsubsection{Werner derivative}

An important generalized class of Werner states is Werner derivative \cite{hiroshima2000local}, which is a mixture of non-maximally pure entangled states and the maximally mixed state. This can be written in the form 
\begin{equation}\label{wd}
    \rho_{WD}=\frac{1-p}{4}\mathds{I}_1\otimes \mathds{I}_2 + p|\psi\rangle\langle \psi |\;,
\end{equation}
where $|\psi\rangle = \sqrt{a}|00\rangle + \sqrt{1-a}|11\rangle$ is the Schmidt decomposition of the state obtained by a nonlocal unitary rotation of the Bell state $|\psi_-\rangle$, and $\frac{1}{2}\le a \le 1$. It is worth noting the difference between the states (\ref{wd}) and (\ref{21}). In the computational basis of two qubits, $\rho_{WD}$ 
takes the following form:
\begin{equation} \label{30}
\rho_{WD}=\frac{1}{4}\begin{pmatrix}
1-p+4ap & 0 & 0 & 4p\sqrt{a(1-a)} \\
0 & 1-p & 0 & 0 \\
0 & 0 & 1-p & 0 \\
4p\sqrt{a(1-a)} & 0 & 0 & 1+3p-4ap \\
\end{pmatrix}\;,
\end{equation}
According to the PT-criterion, the state described by Eq. (\ref{wd}) is entangled if 
\begin{equation}\label{cond_wd}
\frac{1}{2} \le a < \frac{1}{2}\left(1+\frac{1}{2p}\sqrt{(3p-1)(p+1)}\right)\;,
\end{equation}
which further restricts $p$ as $\frac{1}{3}\le p \le 1$. Clearly, for different values of $p$, the parameter $a$ has an upper and a lower bound, such that the state $\rho_{WD}$ parameterized by $a$ becomes entangled. But when using the standard uncertainty relation (\ref{10}), one finds that the state is separable for all $p\in [0,1]$. This can be seen by using $\rho_{WD}^{PT}$ along with the observables $A =\sigma_z\otimes\ \mathds{I}_2$ and $B = \mathds{I}_1\otimes\sigma_z$ in the inequality (\ref{10}), which leads to $p\le 1$. This means, according to (\ref{10}), the state $\rho_{WD}$ should be always separable, which is not the case. We show below, how the inequality (\ref{20}) can successfully detect entanglement in this state.  

To employ the criterion (\ref{20}), we choose the same set of local operators, as above and we obtain the following inequality:
\begin{eqnarray}
&&\left[\frac{3+p-4(2a-1)^{2}p^{2}+D}{2}\right]\nonumber\\
&&\times\left(\frac{1-p-D}{2}\right)\ge \left|\left(\frac{1-p-D}{2}\right)\right|^2\;,
\end{eqnarray}
where 
\begin{equation}
    D=\sqrt{16p^{2}a^{2}-16p^{2}a+(1-p)^{2}}\;.
\end{equation}
We find that the above inequality is violated in the domain when $D$ is imaginary. This happens in the following range of $a$:
\begin{equation}\label{a_range}
    \frac{1}{2}-\frac{1}{4p}\sqrt{(3p-1)(p+1)}\le a \le \frac{1}{2}+\frac{1}{4p}\sqrt{(3p-1)(p+1)}\;.
\end{equation}
The upper limit of $a$ thus matches with the one obtained by directly applying the PT criterion [see Eq. (\ref{cond_wd})]. By definition of the Schmidt decomposition, one further requires $a$ to be real positive, and therefore $p\ge 1/3$ [else, $a$ would be complex; see (\ref{a_range})]. Note that $p$ cannot be greater than unity, as it defines the probability of the state $|\psi\rangle$ in the mixture $\rho_{WD}$. Interestingly, for $p=1/3$, the state $\rho_{WD}$ is entangled only for $a=1/2$. For higher values of $p$, the Werner derivative is entangled for a range of values $a$, including $a=1/2$ (corresponding to the maximally entangled Bell state) and  $a\ne 1/2$ (corresponding to a non-maximally entangled state $|\psi\rangle$).

\subsubsection{An example of pure nonmaximally entangled state}
To further check the criterion (\ref{20}), we next consider a non-maximally pure entangled state $|\psi\rangle$ of the following form:
\begin{equation}\label{31}
  |\psi\rangle=c_0 |00\rangle + c_1|11\rangle \;, 
\end{equation}
where $c_0,c_1$ are the complex coefficients and unequal in magnitude. Note that when $c_0=c_1$, the state becomes one of the Bell states, which are maximally entangled.
For $A =\sigma_z\otimes\sigma_z$ and $B=\sigma_x\otimes\sigma_x$, the inequality (\ref{20}) leads to
\begin{equation}\label{32}
0  \ge  |c_0^{*}c_1+c_0c_1^{*}|^{2}\;,
\end{equation}
which is always violated for any nonzero $c_0$ and $c_1$. This is maximally violated when $c_0=c_1$.

\subsubsection{An example of mixed nonmaximally entangled state}
Finally we consider a non-maximally entangled mixed state $\rho_{new}$ \cite{adhikari2008teleportation}, which is a convex combination of a separable density matrix $\rho_{12}^G={\rm Tr}_3(|GHZ\rangle_{123})$ and an inseparable density matrix $\rho_{12}^W={\rm Tr}_3(|W\rangle_{123})$. Here $|GHZ\rangle_{123}$  and $|W\rangle_{123}$ are the GHZ state and W-state, respectively, of three qubits 1, 2, and 3. The state $\rho_{new}$ can be explicitly written as
\begin{equation}\label{33}
\rho_{new} = (1-p)\rho_{12}^G + p\rho_{12}^W   \;,   
\end{equation}
where $0\le p \le 1$. Note that the Werner state is also a convex sum of a maximally entangled pure state and a maximally mixed state. On the contrary, the state $\rho_{12}^W$ is not a pure state (though entangled) and the $\rho_{12}^G$ is also not maximally mixed (though separable). 

In the computational basis of two qubits, $\rho_{new}$ and $\rho_{new}^{PT}$ take the following forms:
\begin{equation}\label{34}
\rho_{new} =\begin{pmatrix}
\frac{3-p}{6} & 0 & 0 & 0 \\
0 & \frac{p}{3} & \frac{p}{3} & 0 \\
0 & \frac{p}{3} & \frac{p}{3} &0 \\
0 & 0 & 0 & \frac{1-p}{2} \\
\end{pmatrix}
\end{equation}
and
\begin{equation}\label{35a}
\rho_{new}^{PT} =\begin{pmatrix}
\frac{3-p}{6} & 0 & 0 & \frac{p}{3}\\
0 & \frac{p}{3} & 0 & 0 \\
0 & 0 & \frac{p}{3} &0 \\
\frac{p}{3} & 0 & 0 & \frac{1-p}{2} \\
\end{pmatrix}\;.
\end{equation}

According to the PT criterion, that $\rho_{new}$ is entangled for $p >0.708$ can be easily verified by finding the eigenvalues of the $\rho_{new}^{PT}$. But this cannot be revealed by using the Eq. (\ref{10}) with $\rho_{new}^{PT}$ and the following set of local operators: $A =\sigma_z\otimes\ \mathds{I}_2$ and $B =
\mathds{I}_1\otimes\sigma_z$. This leads to the following inequality: $p\ge 0$, which means that, according to (\ref{10}), the state $\rho_{new}$ is separable for all $p$. On the contrary, as we show below, the inequality (\ref{20}) can successfully detect the entanglement in this state, as well.

To evaluate the condition (\ref{20}), we choose the same set of local operators, as above and obtain the following inequality:  
\begin{eqnarray}
&&\left(\frac{-18-2p^{2}+24p+12\sqrt{-p^{2}-12p+9}}{9}\right)\nonumber\\  
&&\times \left(\frac{12-8p-4\sqrt{-p^{2}-12p+9}}{3}\right)\nonumber\\
&&\ge \left|\frac{12-8p-4\sqrt{-p^{2}-12p+9}}{3}\right|^{2}\;.
\end{eqnarray}
We find that this is always violated for $p >0.708$, which correctly matches with the result obtained by directly using the Peres criterion.

\subsection{Discussions}
It is worth noting that the usefulness of the criterion (\ref{20}) becomes more prominent when the state under consideration is mixed in nature. In fact, when $\rho$ is pure, the classical mixing is zero and therefore $I(\rho^{PT},A)=V(\rho^{PT},A)$ and similarly for $B$. So, we have $U(\rho^{PT},A)=V(\rho^{PT},A)$ and $U(\rho^{PT},B)=V(\rho^{PT},B)$. The left hand side of (\ref{20}) becomes the same as that in (\ref{10}). On the right hand side also   $C_{\rho^{PT}}(A,B)$ becomes equal to the covariance $Cov_{\rho^{PT}}(A,B)$ for pure state, which are defined by, for any $\rho$,
\begin{equation}
    Cov_\rho (A,B) = {\rm Tr}(\rho AB) - ({\rm Tr}\rho A)({\rm Tr}\rho B)\;.
\end{equation}
In this way, the criterion (\ref{10}) becomes enough to identify the entanglement in two-qubit pure states. But it fails to identify the entanglement in two-qubit mixed state as we have shown that it is satisfied by all the two-qubit mixed states, considered in this paper. 

Note that if the partial transpose is taken on the operators $A$ and $B$ instead of on $\rho$, the inequality (\ref{20}) reduces to the Schrodinger-Robertson partial transpose (SRPT) inequality \cite{gillet2008multipartite}, as given by
\begin{eqnarray} \label{35}
&&(\Delta A^{PT})^{2}(\Delta B^{PT})^{2} \ge 
\frac{1}{4}|\langle [A,B]^{PT}\rangle|^{2}\nonumber\\
&&+\frac{1}{4}|\langle
\{A,B\}^{PT}\rangle
-2\langle A^{PT}\rangle\langle B^{PT}\rangle|^{2}\;.
\end{eqnarray}
This inequality is able to detect entanglement in any pure entangled state of bipartite and tripartite systems, by experimentally measuring mean values and variances of different observables \cite{gillet2008multipartite}. For mixed states, the above inequality detects entanglement of bipartite Werner states better than the Bell inequalities. On the contrary, the criterion (\ref{20}) cannot be experimentally verified, since it involves term like ${\rm Tr}(\sqrt{\rho}A\sqrt{\rho}A)$, which cannot be measured by usual quantum measurements. However, it is possible to set a nontrivial lower bound. For all $\rho$ and $A$, we have
$\frac{1}{2}{\rm Tr}[\rho,A]^{2}\ge {\rm Tr}[\sqrt{\rho},A]^{2}$\cite{girolami2014observable}.  This implies that
$I(\rho,A)\ge I^{L}(\rho,A)\ge 0$, i.e, the skew information has a non-negative lower bound.
For the spectral decomposition, $\rho=\sum_{i}\lambda_{i}|\phi_{i}\rangle\langle \phi_{i}|$, putting $A_{ij}=\langle \phi_{i}|A|\phi_{j}\rangle$, we have 
$I(\rho,A)=\frac{1}{2}\sum_{ij}(\sqrt{\lambda_i}-\sqrt{\lambda_j})^{2}|A_{ij}|^{2}$, with the lower bound 
$I^{L}(\rho,A)=\frac{1}{4}\sum_{ij}(\lambda_i-\lambda_j)^{2}|A_{ij}|^{2}$. This lower bound is experimentally measurable.\\
\section{Conclusions}
In conclusion, we have formulated a strong entanglement criterion for mixed states. This criterion uses the Peres-Horodecki partial transposition applied on a suitable uncertainty relation. We show by explicit analysis that this criterion can be useful for not only the pure states, but also several generalized forms of mixed states. For example, it can correctly reveal the lower bound of the mixing probability (i.e., $p>1/3$) of the Bell state in the Werner state. Thus it turns out to be stronger than any other known criteria, based on, e.g., the Bell inequality, the uncertainty relation proposed in \cite{guhne2002detection} or the Schrodinger-Robertson inequality. The strength of our criterion lies into the fact that it suitably takes care of the quantum share of the uncertainties (the Wigner-Yanase skew information) and correlations of the relevant observables.  Our criterion reduces to the SRPT inequality for the pure state. 
 \bibliography{References}
 \bibliographystyle{ieeetr}
\end{document}